\documentclass[twocolumn]{aastex63}

\usepackage{natbib}
\usepackage{amsmath}
\usepackage{url}
\usepackage{graphicx}
\usepackage{color}

\newcommand{\kms}{\,km\,s$^{-1}$}

\newcommand{\msun}{M$_{\sun}$}
\def\lsim{\hbox{\rlap{\raise 0.425ex\hbox{$<$}}\lower 0.65ex\hbox{$\sim$}}}
\def\gsim{\hbox{\rlap{\raise 0.425ex\hbox{$>$}}\lower 0.65ex\hbox{$\sim$}}}

\def\arcsec{\hbox{$^{\prime\prime}$}}
\def\arcdeg{\mbox{$^\circ$}}

\shorttitle{SN~2019yvq: [\ion{Ca}{2}] and a Double-Detonation Explosion}
\shortauthors{Siebert, Dimitriadis, Polin, \& Foley}

\graphicspath{{./}{figures/}}

\begin{document}

\title{Strong Calcium Emission Indicates that the Ultraviolet-Flashing Type Ia SN~2019yvq was the Result of a Sub-Chandrasekhar Mass Double-Detonation Explosion}

\correspondingauthor{Matthew~R.~Siebert}
\email{msiebert@ucsc.edu}

\author{Matthew~R.~Siebert}
\affiliation{Department of Astronomy and Astrophysics, University of California, Santa Cruz, CA 95064, USA}

\author{Georgios~Dimitriadis}
\affiliation{Department of Astronomy and Astrophysics, University of California, Santa Cruz, CA 95064, USA}

\author{Abigail~Polin}
\affiliation{The Observatories of the Carnegie Institution for Science, 813 Santa Barbara St., Pasadena, CA 91101, USA}
\affiliation{TAPIR, Walter Burke Institute for Theoretical Physics, 350-17, Caltech, Pasadena, CA 91125, USA}

\author{Ryan~J.~Foley}
\affiliation{Department of Astronomy and Astrophysics, University of California, Santa Cruz, CA 95064, USA}

\begin{abstract}
    We present nebular spectra of the Type Ia supernova (SN~Ia) SN~2019yvq, which had a bright flash of blue and ultraviolet light after exploding, followed by a rise similar to other SNe~Ia.  Although SN~2019yvq displayed several other rare characteristics such as persistent high ejecta velocity near peak brightness, it was not especially peculiar and if the early ``excess'' emission were not observed, it would likely be included in cosmological samples.  The excess flux can be explained by several different physical models linked to the details of the progenitor system and explosion mechanism.  Each has unique predictions for the optically thin emission at late times.  In our nebular spectra, we detect strong [\ion{Ca}{2}] $\lambda\lambda$7291, 7324 and Ca~NIR~triplet emission, consistent with a double-detonation explosion.  We do not detect H, He, or [\ion{O}{1}] emission, predictions for some single-degenerate progenitor systems and violent white dwarf mergers.  The amount of swept-up H or He is $<${}$2.8 \times 10^{-4}$ and $2.4 \times 10^{-4}$~M$_{\sun}$, respectively.  Aside from strong Ca emission, the SN~2019yvq nebular spectrum is similar to those of typical SNe~Ia with the same light-curve shape. Comparing to double-detonation models, we find that the Ca emission is consistent with a model with a total progenitor mass of 1.15~M$_{\sun}$. However, we note that a lower progenitor mass better explains the early light-curve and peak luminosity. The unique properties of SN~2019yvq suggest that thick He-shell double-detonations only account for $1.1^{+2.1}_{-1.1}\%$ of the total ``normal" SN~Ia rate.  SN~2019yvq is one of the best examples yet that multiple progenitor channels appear necessary to reproduce the full diversity of ``normal'' SNe~Ia.
\end{abstract}

\keywords{supernovae}

\section{Introduction}\label{s:intro}

Type Ia supernovae (SNe~Ia) are energetic thermonuclear explosions that have produced roughly half of the iron conent of the local Universe \citep[e.g.,][]{Tinsley80, Matteucci86}, shape and heat the interstellar medium \citep[e.g.,][]{Springel2003}, and are excellent cosmological distance indicators from which we can constrain the nature of dark energy \citep[e.g.,][]{Riess98:lambda, Perlmutter99, Scolnic18:ps1, Jones19}.  Major new facilities such as the Vera C.\ Rubin Observatory and the {\it Nancy Grace Roman Space Telescope} are being designed with SN~Ia observations being a top priority \citep{Spergel15, Hounsell18, Ivezic19}.  Despite their critical importance in element creation, galaxy feedback, and cosmology, we still do not know the precise progenitor system and explosion mechanism for SNe~Ia.

From both theory and observations, we know that SNe~Ia come from C/O white dwarfs (WDs) in binary systems \citep{Hoyle60, Colgate69, Nomoto84:w7, Nugent11, Bloom12}.  The companion star may be another WD \citep[i.e., the double-degenerate or DD scenario;][]{Iben84, Webbink84} or a non-degenerate star \citep[i.e., the single-degenerate or SD scenario;][]{Whelan73, Iben96}.  There is strong observational evidence that DD progenitors are responsible for at least some individual SNe~Ia \citep[e.g.,][]{Li11:11fe, Schaefer12, Kelly14, Jacobson-Galan18}, while some SNe~Ia almost certainly came from SD systems \citep[e.g.,][]{Dilday12, Graham19, Kollmeier19}.  Population studies also have somewhat conflicting results where SD and DD progenitor systems may produce SNe~Ia at roughly similar rates \citep[e.g.,][]{Maoz08, Foley12:csm}.

An additional dimension is the explosion mechanism.  While the explosion must be triggered through mass transfer, this can be done quickly or slowly, with hydrogen or helium, and the explosion can start near the center of the star or at the surface, and the primary WD can vary in mass from about 0.7--1.4~M$_{\sun}$.

Despite the m\'elange of progenitor systems and explosion mechanisms, the near-peak luminosity spectral-energy distributions from multi-dimensional radiative-hydrodynamical explosion simulations appear generally similar to each other and observations.  These predicted observables diverge some for epochs only a few days after explosion.  In particular, some models predict a smooth increase in flux from explosion to peak, while others have ``excess'' flux relative to the smooth models for the first few days after explosion.  In particular, this excess flux can be generated by interaction with a non-degenerate companion \citep[if viewed from a particular position;][]{Kasen10:prog}, interaction with circumstellar material \citep{Raskin13, Piro16}, a violent merger of two WDs \citep{Kromer16}, radioactive $^{56}$Ni in the outer layers of the ejecta \citep{Piro16, Noebauer17}, or a ``double-detonation'' where a surface He layer explosively burns, causing a second explosion in the interior of the WD \citep{Woosley11, Nomoto18, Polin19}.

Wide-field, high-cadence surveys have recently discovered several examples of SNe~Ia with this signature \citep{Marion16, Hosseinzadeh17:17cbv, Dimitriadis19:18oh_k2, Shappee19} and additional peculiar thermonuclear WD SNe with excess flux \citep{Cao15, Jiang17}.  While the different scenarios described above predict different durations, luminosities, and colors, the differences are subtle enough that current data sets cannot adequately distinguish between the scenarios (or the predictions all diverge significantly from the observations).  However, these models predict vastly different observables at late times ($\gtrsim$150~days after explosion).  In particular, interaction models predict strong H or He emission lines \citep{Mattila05, Leonard07, Botyanszki18}, the violent merger should have significant unburned material and thus strong [\ion{O}{1}] lines \citep{Maeda08, Taubenberger09}, while a double-detonation can have incomplete core burning and produce a significant amount of Ca throughout the ejecta leading to strong [\ion{Ca}{2}] lines \citep{Polin20:neb}.  Such analyses were performed for the normal SNe~Ia 2017cbv and 2018oh that had early excess flux, but none of the signatures outlined above were seen \citep{Sand18, Dimitriadis19:18oh_neb, Tucker19}.

While [\ion{O}{1}] and H$\alpha$ emission lines have been detected in nebular spectra of SNe~2010lp \citep{Taubenberger13:10lp} and 2018fhw \citep{Kollmeier19, Vallely19}, respectively, there has not been an unambiguous detection of strong [\ion{Ca}{2}] similar to predictions for models of double-detonations. The [\ion{Ca}{2}] $\lambda\lambda$7291, 7324 doublet overlaps with the 7300~\AA\ emission complex which is often understood to be the blending of [\ion{Fe}{2}] and [\ion{Ni}{2}] emission lines in normal SNe~Ia. This feature appears stronger in low-luminosity SNe~Ia and is likely caused by the additional presence of a [\ion{Ca}{2}] component \citep{Mazzali97, Blondin18}. The detection of [\ion{Ca}{2}] is further complicated by diversity in morphologies observed in the 7300~\AA\ feature. This feature often exhibits multiple peaks which are commonly attributed to different elemental species, however, in some cases studies have suggested that asymmetric ejecta distributions could be the cause of double-peaked nebular features \citep{Dong15,  Mazzali18, Vallely20}. 

Radiative transfer calculations of bare low-mass C/O WD detonations and double-detonations of WDs with thin He shells reproduce many of the photospheric properties of typical and low-luminosity SNe~Ia \citep{Shen18a, Townsley19}. Specifically, the light curves presented in \citet{Shen18a} exhibit a relationship between peak luminosity and decline rate that is in general agreement with the \citet{Phillips93} relation. Their synthetic spectra of normal and low-luminosity SNe~Ia also show similar line ratios and velocities to observed SNe~Ia. However, \citet{Polin19} show that double-detonation explosions from progenitors with thin or thick He shells may produce a subclass of SNe~Ia with distinct properties of velocity, color, and polarization \citep{Cikota19}. They also show that massive He shells are needed in order to produce the early-time ``flux excess" seen in several SNe~Ia. These models also predict a strong component of [\ion{Ca}{2}] emission in the nebular phase \citep{Polin20:neb}.

SN~2019yvq, which had a flash of ultraviolet and blue light a few days after explosion \citep{Miller20}, provides an excellent opportunity to test these theories through its nebular spectrum.  SN~2019yvq is relatively normal, but has some remarkable features in addition to its early light curve.  In particular, it has a relatively low peak luminosity of $M_{g} \approx -18.5$~mag but high ejecta velocities \citep{Miller20}.  Nevertheless, SN~2019yvq is not so obviously distinct from typical SNe~Ia as to be removed from cosmology samples.  At a phase of $152.7$~days after peak luminosity, we obtained a Keck spectrum of SN~2019yvq, which has strong [\ion{Ca}{2}] emission unlike typical SNe~Ia.

We present observations of SN~2019yvq, including the late-time Keck spectrum in Section~\ref{s:obs}.  We compare SN~2019yvq to other SNe~Ia and models in Section~\ref{s:anal}, demonstrating that SN~2019yvq was likely casued by a double-detonation explosion.  We discuss the implications of our observations and conclude in Section~\ref{s:disc}.

Throughout this paper, we adopt the AB magnitude system, unless where noted, and $33.14\pm0.11$ mag as the distance modulus to NGC 4441 (the host galaxy of SN~2019yvq; \citealt{Miller20}).

\section{Observations \& Data Reduction}\label{s:obs}

We obtained two optical spectra of SN~2019yvq on 2020 Jun 17 UT, with the Low Resolution Imaging Spectrometer \citep[LRIS;][]{Oke95}, mounted on the 10-meter Keck~I telescope at the W.\ M.\ Keck Observatory. At that date, the SN was $\sim$153 rest-frame days past peak brightness \citep[2020 Jan 15.25 UT;][]{Miller20}. We observed SN~2019yvq with a low-resolution setting (1800 and 1430~s blue channel exposures with the B600/4000 grism and two 525~s red channel exposures with the R400/8500 grating, with pixel scales of 0.63 and 1.16~\AA/pixel, respectively) and a high-resolution setting (two 825~s red channel exposures with the R1200/7500 grating, with a pixel scale of 0.4~\AA/pixel).  We used the 1.0\arcsec-wide slit and the D560 dichroic for all observations, and oriented the slit to include the host-galaxy nucleus.  The atmospheric dispersion corrector unit was deployed. The low-resolution spectrum covers 3,400 -- 10,056~\AA, while the high-resolution spectrum covers 6,200 -- 7,800~\AA, including H$\alpha$, \ion{He}{1} $\lambda$6678, [\ion{O}{1}] $\lambda\lambda$6300, 6364 and the 7300~\AA\ line complex, the primary focus of our current analysis.  All data were reduced using standard \textsc{iraf}\footnote{IRAF is distributed by the National Optical Astronomy Observatory, which is operated by the Association of Universities for Research in Astronomy (AURA) under a cooperative agreement with the National Science Foundation.} and python routines for bias/overscan corrections, flat fielding, flux calibration and telluric lines removal, using spectro-photometric standard star spectra, obtained the same night \citep{Silverman12:bsnip}

\begin{figure*}
\begin{center}
    \includegraphics[width=6.4in]{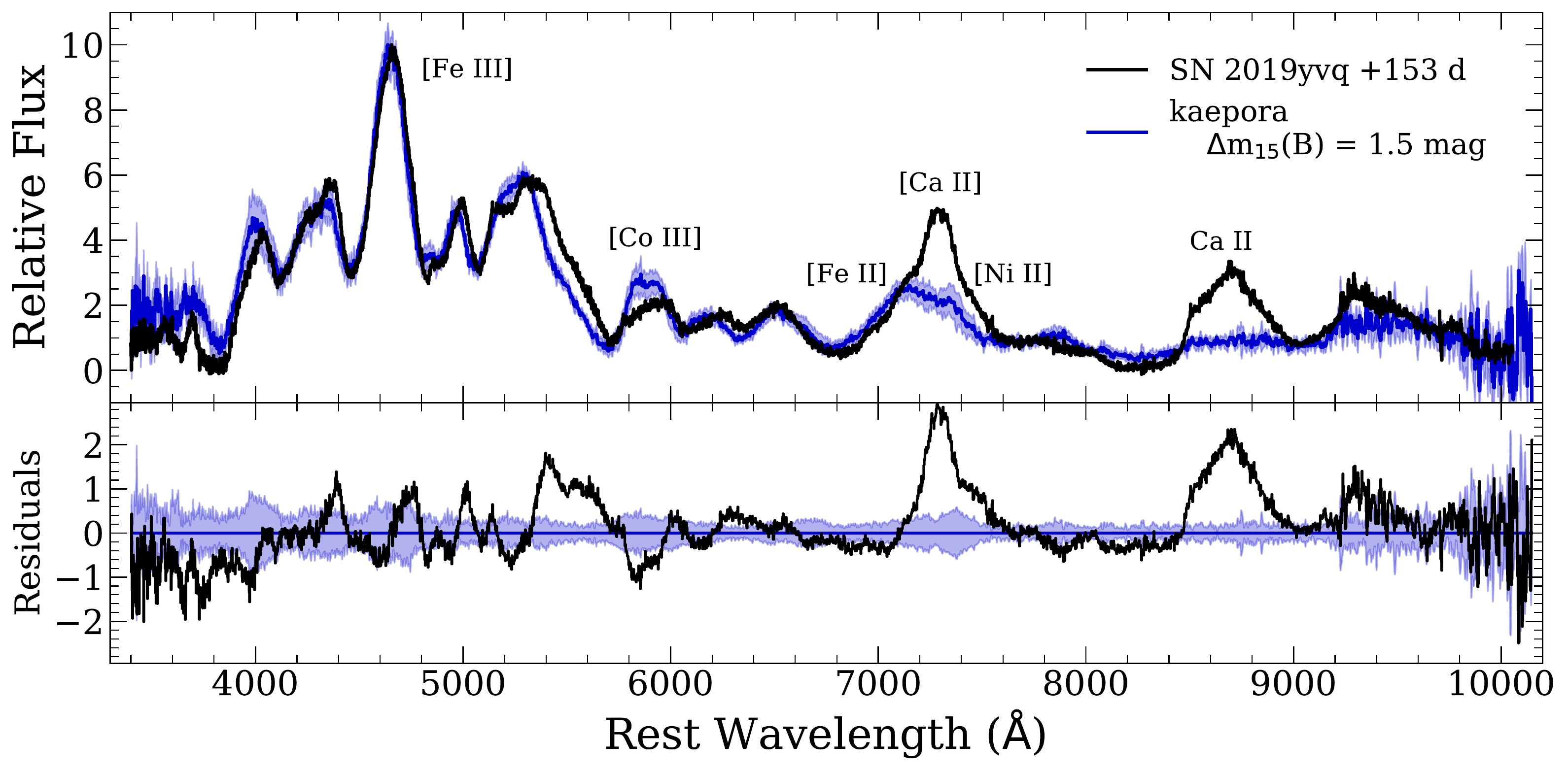}
\caption{{\it Top Panel}: Spectrum of SN~2019yvq (black) observed 153 rest-frame days after peak brightness.  The kaepora $\Delta m_{15} (B) = 1.5$~mag (the same decline rate as SN~2019yvq) composite spectrum \citep{Siebert19} is also displayed (blue) along with the $1{-}\sigma$ scatter of the spectra used to produce the composite spectrum.  {\it Bottom Panel}: Residual spectrum of SN~2019yvq relative to the kaepora comparison spectrum.}\label{fig:compare_kaepora}
\end{center}
\end{figure*}

We present the low-resolution spectrum in \autoref{fig:compare_kaepora}.  The high-resolution spectrum is nearly identical other than its resolution and limited wavelength range.

The nebular spectrum of SN~2019yvq is generally similar to those of other SNe~Ia at a similar epoch, including strong line emission from forbidden singly- and doubly-ionized Fe-group elements.  At this epoch, the spectrum is likely still evolving, but the lack of obvious P-Cygni profiles indicates that the ejecta are mostly or completely optically thin.

Unlike other ``normal'' SNe~Ia, SN~2019yvq has clear and strong [\ion{Ca}{2}] $\lambda\lambda$7291, 7324 and \ion{Ca}{2} NIR triplet emission.  We discuss these feature, the connection to a double-detonation explosion, and the lack of signatures from other progenitor channels in the following sections.

\section{Analysis}\label{s:anal}

\subsection{Photometric comparisons}

SN~2019yvq had a lower peak luminosity than typical SNe~Ia \citep[$M_{g, {\rm ~peak}} \approx -18.5$~mag;][]{Miller20}.  \citet{Miller20} measured a corresponding, relatively fast $g$-band ($\Delta m_{15} (g) = 1.3$~mag).  Most historical SNe~Ia lack a $\Delta m_{15} (g)$ measurement, and so it is difficult to make direct comparisons with other SNe~Ia with similar light-curve shape.  \citet{Miller20} used the \citet{Yao19} relationship between $\Delta m_{15} (g)$ and $\Delta m_{15} (B)$ to estimate $\Delta m_{15} (B) \gtrsim 1.6$~mag for SN~2019yvq.  This analysis was limited by the lack of fast-declining SNe~Ia in the \citet{Yao19} sample, preventing a precise measurement.

Using a sample of SNe~Ia with both $g$ and $B$ light curves \citep{Folatelli13}, we select a subset of five SNe~Ia with similar $g$-band light curves.  These SNe~Ia have an average $\Delta m_{15} (B) = 1.54$~mag with an RMS of $0.07$~mag, consistent with the \citet{Miller20} estimate.  We use our derived estimate as the $B$-band decline rate for SN~2019yvq.

\subsection{Spectroscopic comparisons}

\begin{figure*}
\begin{center}
    \includegraphics[width=6.1in]{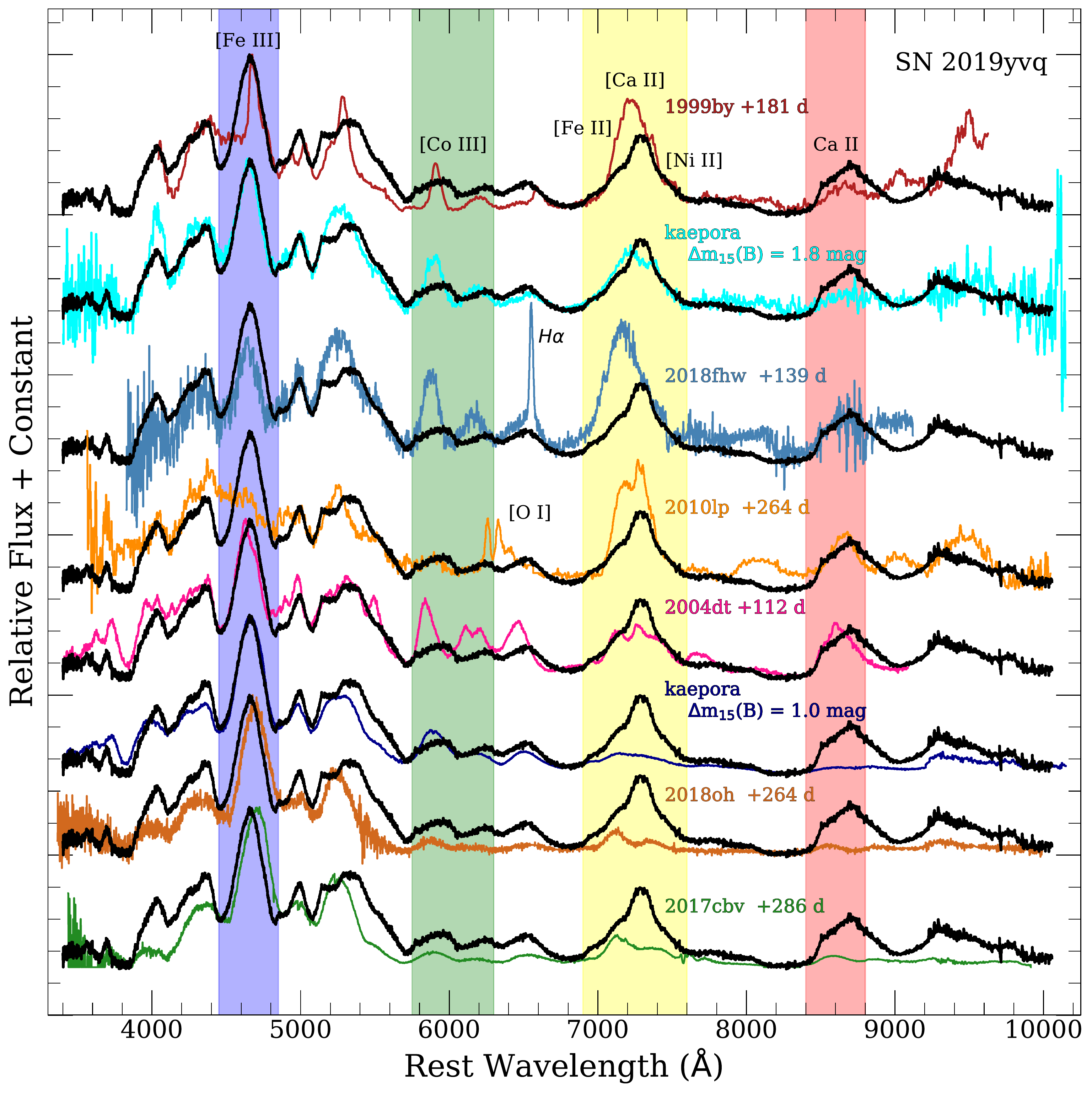}
\caption{Optical spectrum of SN~2019yvq (black curve) at +153~days after peak brightness compared to those of other SNe~Ia at similar phases. From top to bottom we compare to scaled nebular spectra of the SN~1991bg-like SN~1999by (red); the kaepora composite spectrum with $\Delta m_{15} (B) = 1.8$~mag (cyan); SN~2018fhw, which had late-time H$\alpha$ emission (blue); SN~2010lp, a peculiar SN~2002es-like SN which had nebular [\ion{O}{1}] emission (orange), we have clipped emission lines from the host galaxy for better visualization; the high-polarization and peculiar SN~2004dt (fucsia); the kaepora composite spectrum with $\Delta m_{15} (B) = 1.0$~mag (dark blue); SN~2018oh, which had an early-time flux excess (dark orange); and SN~2017cbv, which also had an early-time flux excess (green).  Several spectral regions are highlighted: [\ion{Fe}{3}] $\lambda 4701$ (blue); [\ion{Co}{3}] $\lambda 5888$; the feature at 7300~\AA\ complex which includes possible contributions from [\ion{Fe}{2}] $\lambda 7155$, [\ion{Ni}{2}] $\lambda 7378$, and [\ion{Ca}{2}] $\lambda \lambda 7291$, 7324 (yellow); and the \ion{Ca}{2} NIR triplet (red).}\label{fig:spec_compare}
\end{center}
\end{figure*}

Using the methods of \citet{Siebert19}, we generate a composite spectrum using kaepora\footnote{\url{https://msiebert1.github.io/kaepora/}} to best match the phase and decline rate of SN~2019yvq and compare in \autoref{fig:compare_kaepora}. Aside from the clear Ca emission, SN~2019yvq has remarkably similar line shifts, line widths, and relative feature strengths to those in the kaepora $\Delta m_{15} (B) = 1.5$~mag composite spectrum.

While the optical spectrum of SN~2019yvq at +153~days is generally similar to other SN~Ia spectra at similar epochs, the morphology of the 7300~\AA\ line complex is unique compared to all other known SN~Ia nebular spectra.  In \autoref{fig:spec_compare} we compare this spectrum to a diverse set of SN~Ia nebular spectra.  The SN~2019yvq spectrum is similar to those of typical SNe~Ia and composite spectra in regions without Ca emission.  While some peculiar SNe~Ia appear more similar to SN~2019yvq in wavelength regions corresponding to Ca emission, their spectra are less similar at other wavelengths.

Examining the peculiar SNe~Ia in more detail, we highlight similarities and differences with SN~2019yvq.  \autoref{fig:spec_compare} displays spectra from  SNe~1999by ($M_{B, {\rm ~peak}} = -17.2$~mag; \citealt{Garnavich04}), a low-luminosity SN~1991bg-like SN~Ia; 2010lp ($M_{B, {\rm ~peak}} = -17.7$~mag; \citealt{Kromer13}; Pignata et al., in preparation), a peculiar SN~2002es-like \citep{Ganeshalingam12} SN~Ia that had strong [\ion{O}{1}] $\lambda \lambda$6300, 6364 emission in its late-time spectrum \citep{Taubenberger13} indicating significant unburned material; and 2018fhw ($M_{B, {\rm ~peak}} = -17.7$~mag; \citealt{Kollmeier19}), which had strong H$\alpha$ emission in its late-time spectrum indicating circumstellar interaction \citep{Kollmeier19, Vallely19}. The nebular spectra of SNe~1999by and 2018fhw are very similar overall, except for the strong H$\alpha$ emission seen for SN~2018fhw.  These spectra show the general trends seen in other SN~1991bg-like nebular spectra of narrower features, stronger [\ion{Co}{3}] emission relative to [\ion{Fe}{3}], and a stronger 7300~\AA\ emission complex. SN~2010lp is somewhat different with very weak (perhaps absent) [\ion{Fe}{3}] emission, but shares other characteristics.  Other than the strong emission near 7300~\AA, the SN~2019yvq spectrum does not have the distinct properties of low-luminosity SN~Ia spectra, including the peculiar SNe~2010lp and 2018fhw. Of these comparison spectra, the kaepora composite spectrum with $\Delta m_{15} (B) = 1.8$~mag best reproduces the ratio of the 7300~\AA\ line complex to the [\ion{Fe}{3}] peak, and it is possible that the SNe contributing to the composite spectrum contain a similar contribution from [\ion{Ca}{2}] as for SN~2019yvq.

Alternatively, \autoref{fig:spec_compare} also compares the spectrum of SN~2019yvq to higher-luminosity SNe~Ia, including some relatively peculiar SNe.  We compare to SNe~2004dt, a high-velocity and high-polarization SN~Ia \citep{Wang06,Altavilla07} that is an outlier when comparing its peak-light velocity gradient and nebular-line velocity shifts \citep{Maeda10:asym}; 2017cbv, which had an early blue flux excess days after explosion \citep{Hosseinzadeh17:17cbv}; and 2018oh, which also had a distinct flux excess at early times \citep{Dimitriadis19:18oh_k2, Shappee19}.  Except for the Ca features, the SN~2019yvq spectrum is similar to that of these comparison SNe (although SN~2004dt shows more differences, which may be caused by its relatively early phase).  We also compare the SN~2019yvq spectrum to a kaepora composite spectrum with $\Delta m_{15} (B) = 1.0$~mag showing striking similarity except for the Ca emission. We note that the flux at ${\sim}5500$~\AA\ appears to be strongly correlated with phase. SN~2019yvq shows the best agreement in this wavelength range with SNe~2018fhw and SN~2004dt, which have phases of 139 and 112~days, respectively, significantly earlier than many of the other comparison spectra.

The width and relative strength of [\ion{Fe}{3}] $\lambda 4701$ in SN~2019yvq is most similar to SNe~2004dt, 2011fe, and 2017cbv. Both SN~2019yvq and SN~2017cbv exhibited early blue bumps in their light-curves, however, only SN~2019yvq shows prominent \ion{Ca}{2} features. Of the SNe displayed, only SNe~2004dt, 2010lp, and 2019yvq show prominent \ion{Ca}{2} near-infrared triplet emission. 

\begin{figure*}
\begin{center}
    \includegraphics[width=6.1in]{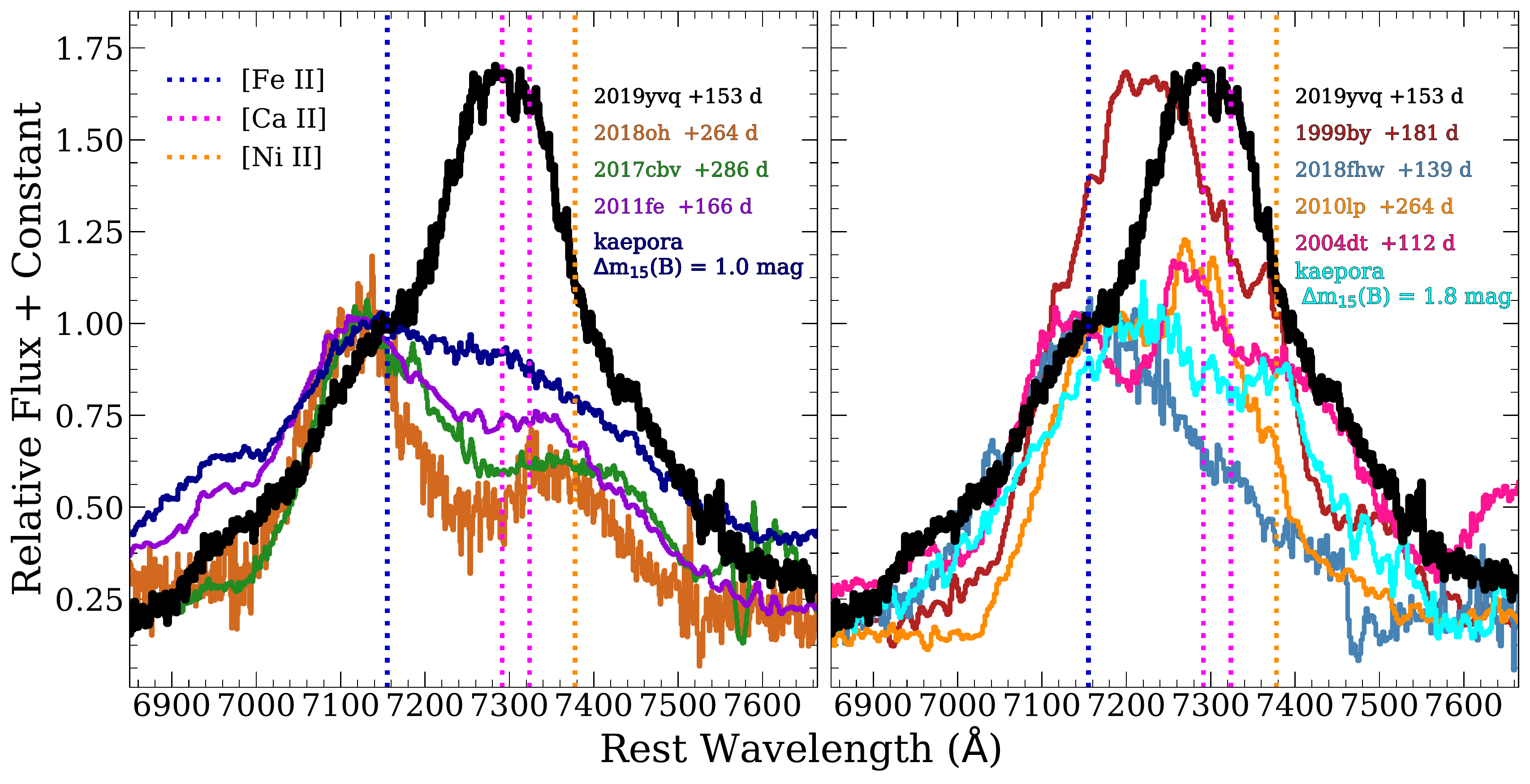}
\caption{(\textit{left panel}): Comparison of the 7300~\AA\ line complex of higher-luminosity SNe (SNe~2011fe, 2017cbv, 2018oh, and the kaepora $\Delta m_{15} (B) = 1.0$~mag  composite spectrum). These spectra have been scaled such that their peak [\ion{Fe}{2}] flux match the the peak of the [\ion{Fe}{2}] component in SN~2019yvq. (\textit{right panel}):  Comparison of the 7300~\AA\ line complex of SNe that may have strong [\ion{Ca}{2}] components (SN~1999by, SN~2018fhw, SN~2010lp, and SN~2004dt). The 7300\AA\ line complex in SN~2018fhw and SN~1999by is relatively dominant in comparison to [\ion{Fe}{3}] as shown in \autoref{fig:spec_compare}. Thus these spectra have been scaled such that their peak [\ion{Ca}{2}] flux matches the the peak of the [\ion{Ca}{2}] component in SN~2019yvq and SN~2004dt and SN~2010lp have been scaled to match the [\ion{Fe}{2}] emission. The rest wavelengths of prominent \ion{Fe}{2}, \ion{Ca}{2}, and \ion{Ni}{2} are displayed as vertical dashed lines in both panels. }\label{fig:spec_7300}
\end{center}
\end{figure*}

We examine the 7300~\AA\ emission feature in detail in \autoref{fig:spec_7300}.  The spectra of the higher-luminosity comparison SNe (SNe~2011fe, 2017cbv, 2018oh, and the kaepora $\Delta m_{15} (B) = 1.0$~mag  composite spectrum) are distinct from that of SN~2019yvq with the comparison spectra having obvious [\ion{Fe}{2}] and [\ion{Ni}{2}], but lacking significant [\ion{Ca}{2}] emission.  This is in contrast to SN~2019yvq, which has strong [\ion{Ca}{2}] emission in addition to the [\ion{Fe}{2}] and [\ion{Ni}{2}] emission.

In \autoref{fig:spec_7300}, we also display the 7300~\AA\ emission feature, comparing SN~2019yvq to the lower-luminosity and peculiar SNe~Ia from \autoref{fig:spec_compare}.  Each of these SNe has possible [\ion{Ca}{2}] emission, but all comparison spectra are still distinct from the SN~2019yvq spectrum.  In particular, SNe~1999by and 2018fhw have strong emission peaking around 7220 and 7160~\AA, respectively, much bluer than SN~2019yvq, which peaks at 7287~\AA.  While it is possible SNe~1999by and 2018fhw have strong [\ion{Ca}{2}] $\lambda \lambda 7291$, 7324 emission, the emitting material would be blueshifted by $-3500$ and $-5900$~\kms, respectively, which would be some of the highest velocity shifts seen for a SN~Ia \citep{Maeda10:neb,Maguire18}, and is inconsistent with shifts from other spectral features.  Instead, it is more likely that there is significant contribution from [\ion{Fe}{2}] $\lambda$7155.

SNe~2004dt and 2010lp have more obvious [\ion{Ca}{2}] $\lambda \lambda 7291$, 7324 emission with peaks at $\sim$7290~\AA, corresponding to velocity shifts of $-1400$ and $-1100$~\kms, respectively.  However, SN~2019yvq has significantly stronger [\ion{Ca}{2}] emission relative to [\ion{Fe}{2}] and [\ion{Ni}{2}] than SNe~2004dt and 2010lp.  We also caution that SN~2004dt has significant velocity offsets and polarization, and it is possible that the emission peak for that particular spectrum is caused by an asymmetric and kinematically extreme ejecta distribution.

There is a general trend between peak luminosity and  the 7300~\AA\ profile shape with lower-luminosity SNe~Ia having stronger emission at these wavelengths relative to other features \citep[see \autoref{fig:spec_compare};][]{Polin20:neb}.  SN~2019yvq conforms to this trend.  However, no other SN~Ia is so sufficiently dominated by [\ion{Ca}{2}] emission at these wavelengths.

SN~2019yvq has a similar [\ion{Ca}{2}]/[\ion{Fe}{2}] strength to some SNe~Iax \citep{Foley16:iax}, albeit the lines have much larger velocity widths for SN~2019yvq.  Other (presumably) white-dwarf SNe such as Ca-rich SNe \citep[e.g.,][]{Perets10:05e} and SN~2016hnk \citep{Jacobson-Galan20} also have strong [\ion{Ca}{2}] emission at late times, but are generally different from SN~2019yvq and SNe~Ia in most regards (e.g., peak luminosity, light-curve behavior, He abundance).

The blue and red slopes of the Ca-deficient spectra in this wavelength range seem to agree well with the shoulders of the line complex of SN~2019yvq. Additionally, the spectrum of SN~2011fe and the kaepora composite spectrum with $\Delta m_{15} (B) = 1.0$~mag show some evidence for excess emission from 6900-7050~\AA. These spectra have phases of +166 days and +140 days, respectively, which are very similar to the spectrum of SN~2019yvq at +153 days. Therefore, this could be a feature that is more likely to be observed at early times. All of the Ca-deficient spectra have [\ion{Fe}{2}] emission components that appear blueshifted relative to SN~2019yvq.

\subsection{Fitting the 7300~\AA\ Line Complex}

The complicated morphology of the 7300~\AA\ line complex allows us to decompose it into emission from different species. Doing this, we can examine the contribution from [\ion{Ca}{2}].

We fit this feature using the following methodology. First, we smooth the spectrum with a 15~\AA\ scale and choose continuum points on the red and blue sides of the feature. We divide by this linear continuum and use Gaussian profiles to approximate the forbidden line emission from [\ion{Fe}{2}], [\ion{Ca}{2}], and [\ion{Ni}{2}]. For this analysis we assume that the following lines dominate this fitting region:  [\ion{Fe}{2}] (7155, 7172, 7388, 7453~\AA), [\ion{Ca}{2}] (7291, 7324~\AA), and [\ion{Ni}{2}] (7378, 7412\AA). We used the rest wavelengths and transition probabilities from the NIST Atomic Spectra Database\footnote{\url{https://www.nist.gov/pml/atomic-spectra-database}}. The strengths of lines for each element are defined relative to the strongest line, which is a single free parameter for each species. We assume that lines produced by unique ionization states of each element are produced in the same regions of the ejecta, and we therefore require that the velocity offsets and widths relative to the rest-frame wavelength of each line be the same for lines coming from the same species. Since [\ion{Fe}{2}] and [\ion{Ni}{2}] emission are likely produced in the same region of the ejecta \citep{Maeda10:neb}, we fit for a single [\ion{Fe}{2}] and [\ion{Ni}{2}] velocity offset. Thus, we fit for a total of 8 parameters: the emission strength, velocity offset ([\ion{Fe}{2}] and [\ion{Ni}{2}], and [\ion{Ca}{2}]), and velocity width of each species. To estimate uncertainties for these parameters, we performed a simple Monte-Carlo algorithm to vary the blue and red continuum points randomly by up to 100~\AA\ and repeating the analysis 

\begin{figure}
\begin{center}
    \includegraphics[width=3.2in]{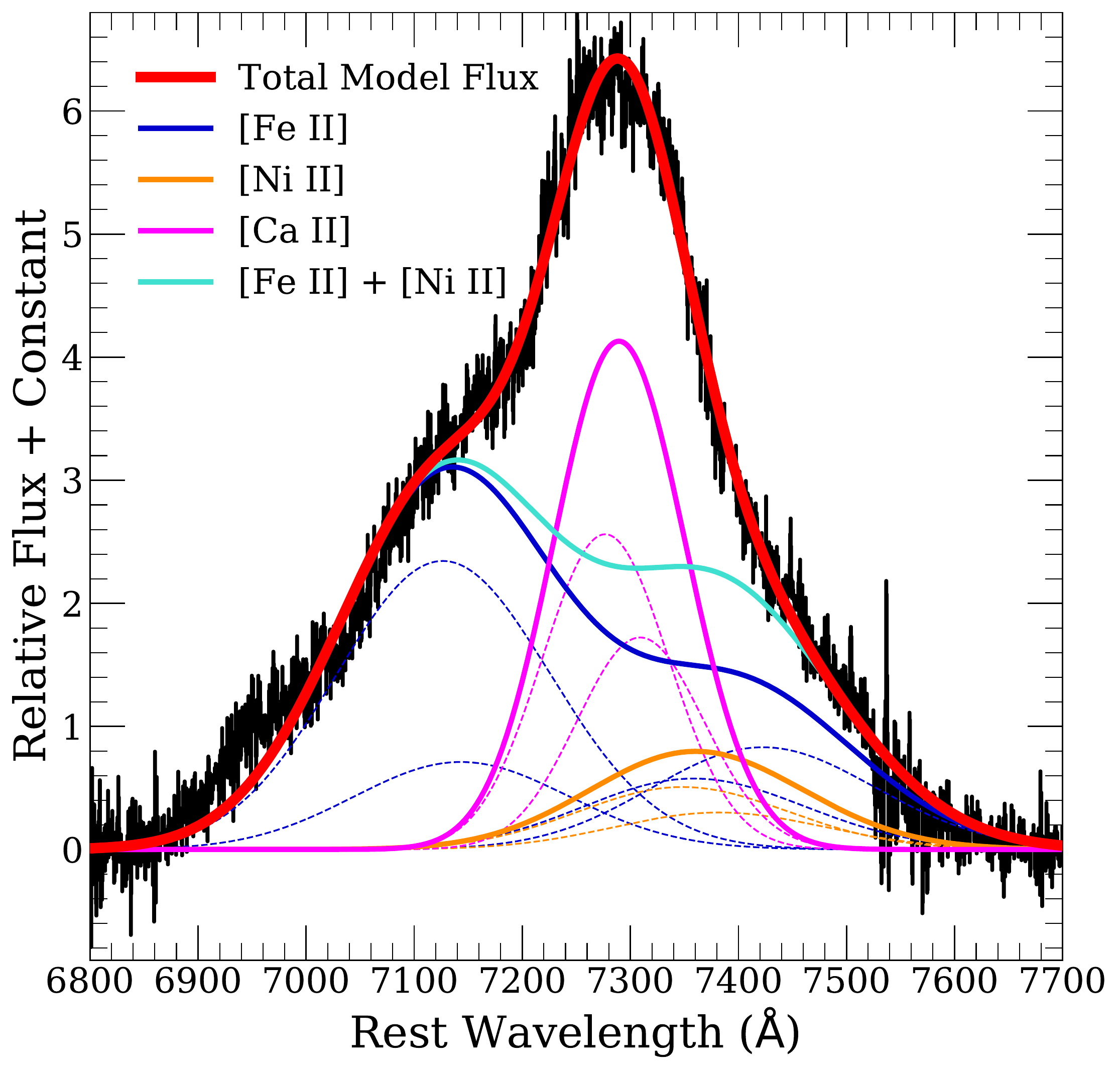}
\caption{Multiple Gaussian-component fit (red) to the 7300~\AA\ line complex in the LRIS high-resolution nebular spectrum of SN~2019yvq (black). Emission from the [\ion{Fe}{2}], [\ion{Ni}{2}], and [\ion{Ca}{2}] and shown as solid blue, teal, and magenta curves, respectively, while dotted lines represent the emission from individual line transitions. A strong [\ion{Ca}{2}] emission component is needed to reproduce explain the emission seen for SN~2019yvq.}\label{fig:spec_fit}
\end{center}
\end{figure}

\autoref{fig:spec_fit} displays the best-fit Gaussian component model to the 7300~\AA\ line complex.  The best-fit values of velocity offsets and widths are presented in \autoref{tab:fit}.  This simple model matches the line profile extremely well.

\begin{deluxetable}{ccc}
\tablehead{
\tablewidth{0pt}
 \colhead{Species} & \colhead{Velocity Offset (km s$^{-1}$)} & \colhead{Width (km s$^{-1}$)}
}

\tablecaption{Parameters for Multiple-Gaussian Decomposition of the 7300~\AA\ line Profile\label{tab:fit}}

\startdata
 \ion{Fe}{2} & $-1210 \pm 90$  & $4170 \pm 70$\\
 \ion{Ni}{2} & $-1210 \pm 90$  & $3960 \pm 20$\\
 \ion{Ca}{2} & $-600 \pm 90$  & $2400 \pm 40$
\enddata

\end{deluxetable}

To match the data and in particular the peak of the emission, a strong [\ion{Ca}{2}] component is necessary. Since Fe-group elements are expected to be produced in similar regions of the ejecta, it is reassuring that our fit produces similar velocity widths for [\ion{Fe}{2}] and [\ion{Ni}{2}] ($4400 \pm 100$ and $3910 \pm 40$~\kms, respectively). Additionally, the relative strength of [\ion{Fe}{2}] to [\ion{Ni}{2}] is consistent with other fits in the literature that use similar methods to fit this feature \citep{Maguire18}. The sum of the the [\ion{Fe}{2}] and [\ion{Ni}{2}] components result in a profile that is qualitatively similar to 7300~\AA\ line profiles of typical SNe~Ia at similar epochs. In particular, the kaepora $\Delta m_{15} (B) = 1.0$~mag composite spectrum, and the nebular spectrum of SN~2011fe (\autoref{fig:spec_7300}, left panel, blue curve and purple curve, respectively) have the most similar morphology to our [\ion{Fe}{2}] + [\ion{Ni}{2}] component. Similar to the nebular spectrum of SN~2019yvq at +153 days, the kaepora $\Delta m_{15} (B) = 1.0$~mag composite spectrum has a effective phase of +138~days, and the SN~2011fe spectrum is at +166~days.

All three species (\ion{Fe}{2}, \ion{Ni}{2}, and \ion{Ca}{2}) are blueshifted relative to the rest frame. \citet{Maeda10:neb} found that nebular line shifts are correlated with velocity gradient. \citet{Blondin12}, \citet{Silverman13:late}, and \citet{Maguire18} supported this result by showing that high-velocity SNe~Ia are more likely to have redshifted nebular lines. Given that SN~2019yvq exhibited a high ejecta velocity at peak brightness of about $-15{,}000$~\kms\ \citep{Miller20}, a blueshifted nebular velocity is atypical --- similar to how its red intrinsic color at peak is atypical for this high velocity \citep{Foley11:vel, Foley11:vgrad}.

The full complex of SN~2019yvq cannot be fit without [\ion{Ca}{2}] emission, unlike what is seen for most SNe~Ia \citep[e.g.,][]{Maguire18, Flors20}.  Furthermore, the additional presence of strong Ca II near-infrared triplet emission provides more evidence that the strong component of the 7300~\AA\ line profile is caused by [\ion{Ca}{2}]. Other low-luminosity SNe~Ia such as SNe~1991bg and 1999by have strong emission in this region reminiscent of SN~2019yvq and this is often attributed to [\ion{Ca}{2}] \citep[e.g.,][]{Filippenko92:91bg, Turatto96}, it is not easily reproduced by (only) [\ion{Ca}{2}] \citep{Mazzali97}.  However, \citet{Blondin18} presented model spectra of SN~1999by where [\ion{Ca}{2}] dominated this feature with an additional strong component from [\ion{Ar}{3}] $\lambda$7136. Currently there is no unambiguous, dominant [\ion{Ca}{2}] emission in a SN~1991bg-like SN~Ia.  Moreover, [\ion{Ar}{3}] $\lambda$7136 is not a strong line in the SN~2019yvq spectrum.

Alternatively, [\ion{Ca}{2}] has been clearly detected in several peculiar SNe~Ia and SNe~Iax \citep{Taubenberger13:10lp, Foley16:iax, Galbany19, Jacobson-Galan20}, indicating that this feature is detectable at the expected wavelength under the correct physical conditions.  SN~2016hnk and SNe~Iax are both connected to He burning \citep{Foley13:iax, Foley16:iax, Jacobson-Galan19, Jacobson-Galan20}, perhaps further indicating SN~2019yvq is the result of a double-detonation.

\citet{Wilk20} argued that the presence of [\ion{Ca}{2}] blended with the 7300~\AA\ line complex allows for the constraint of the ionization ratio. They suggest that for $N({\rm Fe}^+)/N({\rm Ca}^+) \geq 50$, [\ion{Fe}{2}] is expected to dominate this feature yet for $N({\rm Fe}^+)/N({\rm Ca}^+) \leq 100$ prominent [\ion{Ca}{2}] blending is expected. Thus, the unambiguous detection of both [\ion{Fe}{2}] and [\ion{Ca}{2}] in SN~2019yvq may allow us to constrain the number ratio of ionized Fe to Ca, $N({\rm Fe}^+)/N({\rm Ca}^+)$, to between 50 and 100. \citet{Wilk20} also found that significant clumping of the ejecta is a natural way to decrease ionization resulting in stronger Ca emission as [\ion{Ca}{2}] becomes the dominant cooling for regions rich in IMEs.

\subsection{Mass Limits For Swept-up Circumstellar Material} \label{sec:mass_limits_hydrogen_helium}

\begin{figure}
\begin{center}
  \includegraphics[width=.98\linewidth]{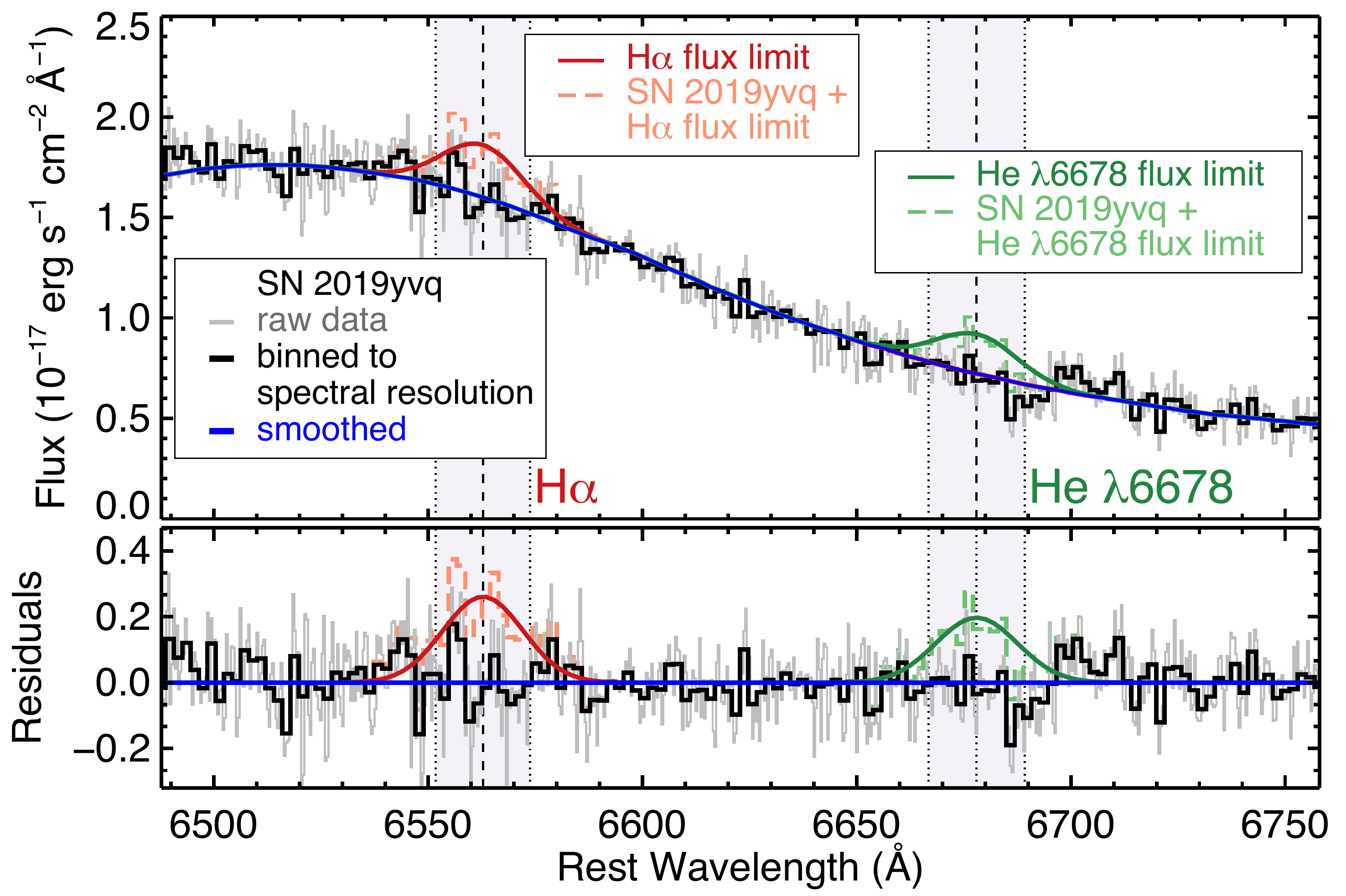}
\caption{The LRIS high resolution spectrum of SN\,2019yvq, at the spectral region of H$\alpha$ and \ion{He}{1} $\lambda6678$. The solid gray line corresponds to the raw data, and the solid black line to the raw data binned to the spectral resolution. The underlying continuum is shown as a solid blue line. The gray-shaded region corresponds the $\pm$22~\AA\ (1000~\kms) region around the rest wavelength of each line. Solid red and green lines represent the artificially-inserted H$\alpha$ and \ion{He}{1} $\lambda6678$ features, corresponding to our 3-$\sigma$ detection limit above the smoothed continuum, with the dashed red and green lines showing how these features would appear in our spectrum. On the bottom panel, we additionally show the residuals relative to the continuum.  }\label{f:hydrogen_helium_limit}
\end{center}
\end{figure}

A visual inspection of the late-time SN~2019yvq spectra shows no obvious hydrogen or helium emission at the redshift of the SN. We can, alternatively, constrain the amount of swept-up material from a potential companion to the exploding WD, following the procedure described in several SNe-Ia nebular studies \citep{Mattila05, Leonard07, Shappee13, Maguire16, Graham17, Sand18,  Dimitriadis19:18oh_neb, Tucker20:nebular} as follows: Firstly, we estimate the brightness of SN~2019yvq at +200~days past explosion in order to compare our data with the models from \citet{Botyanszki18}. While our spectrum was taken at $\sim$170~days from explosion, the late-time spectral features of SNe Ia do not change significantly between these epochs, thus the general spectral shape of SN~2019yvq at +200~days should be similar. We use the public $g_{\rm ZTF}$ and $r_{\rm ZTF}$ photometry after +60~days from maximum, when the SN is in the radioactive cobalt decay regime, correct for MW and host extinction (using the values from \citealt{Miller20}) and linearly fit, estimating $g_{+200 {\rm ~d}} = 19.67 \pm 0.10$~mag and $r_{+200 {\rm ~d}} = 20.77 \pm 0.14$~mag. Finally, we warp our late-time spectrum to match these estimated photometric colors.

To determine the mass limits, we follow the procedure outlined by \citet{Dimitriadis19:18oh_neb}.  Briefly, using the flux-calibrated, extinction- and redshift-corrected spectrum, we bin to the spectral resolution. Our high-resolution spectrum has FWHM spectral resolution of $\sim$1.95~\AA\ (as determined from isolated night-sky lines).  The pixel scale is $\sim$0.4~\AA, and thus not limiting the resolution.  We determine the continuum by smoothing on a 195~\AA\ scale.  Comparing the smoothed spectrum to the unsmoothed version, we do not detect any significant emission features expected from the interaction scenario.  Approximating possible emission features as Gaussians with FWHMs of 1000~\kms, we determine the 3-$\sigma$ flux limit.  Using the luminosity distance from \citealt{Miller20}, we estimate the H$\alpha$ and \ion{He}{1} $\lambda\lambda6678$ luminosity limits (at 200 days) to be 1.33 and $1.01 \times 10^{37}$~erg~s$^{-1}$, respectively. Using Equation~1 of \citet{Botyanszki18}, we convert these luminosity limits to mass limits, and we determine that SN\,2019yvq had stripped hydrogen and helium mass of $<${}$4.0 \times 10^{-4}$ and $3.4 \times 10^{-4}$~M$_{\sun}$, respectively. These results are displayed in Figure~\ref{f:hydrogen_helium_limit}.

\subsection{Comparison to Double-Detonation Model}

\begin{figure*}
\begin{center}
    \includegraphics[width=.49\linewidth]{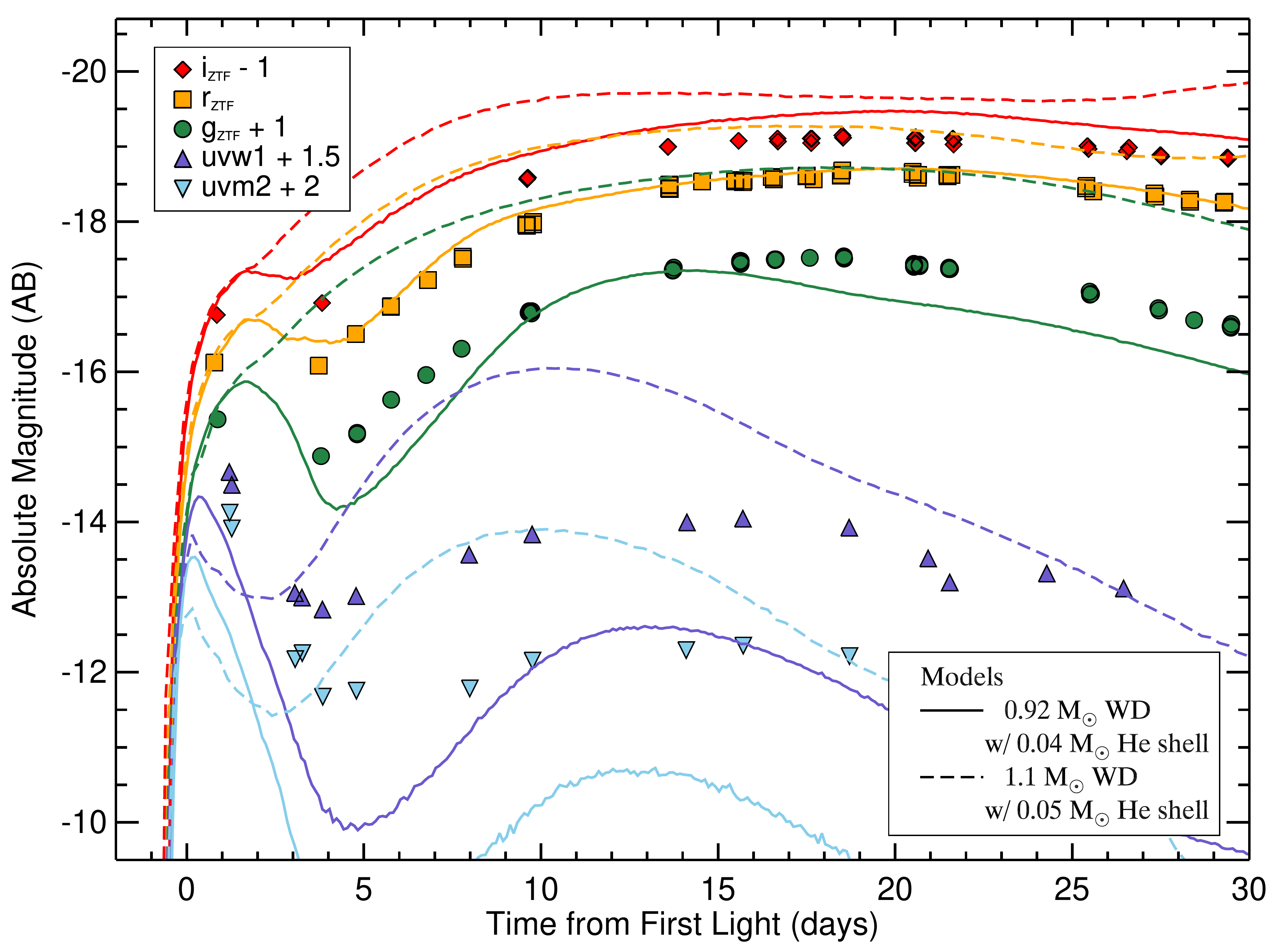}
    \includegraphics[width=.49\linewidth]{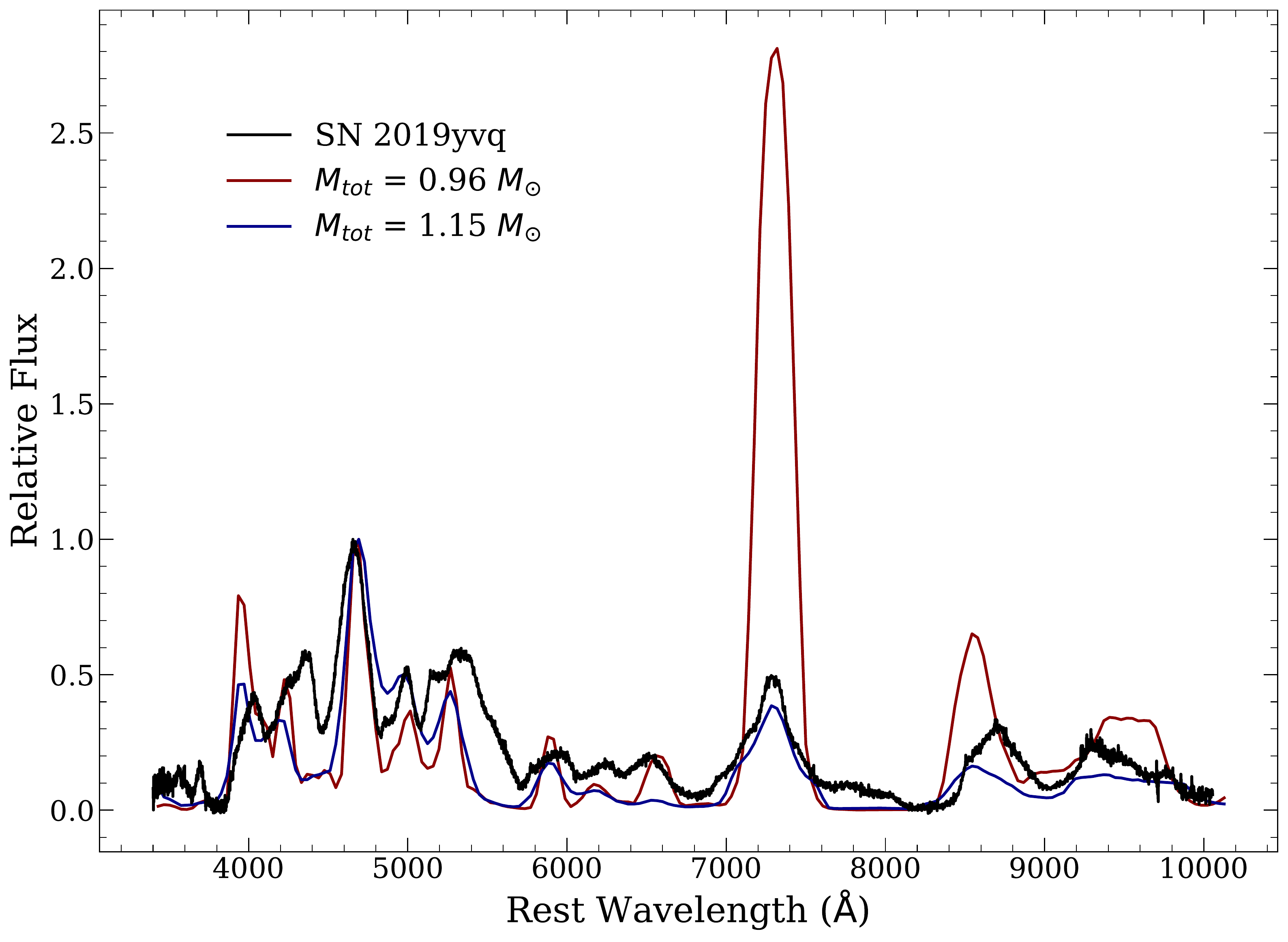}
\caption{SN~2019yvq light curves (left) and nebular spectrum (right) compared with double-detonation models.  
The light curves of the \citet{Miller20} model (a 0.92~\msun\ WD with 0.04~\msun\ He on its surface) are displayed as solid lines, while the spectrum is displayed as a red curve.  The light curves of an additional model that is well matched to the nebular spectrum (a 1.1~\msun\ WD with 0.05~\msun\ He on its surface) is displayed as dashed curves, and its nebular spectrum is a blue curve.  The \citet{Miller20} model has nebular [\ion{Ca}{2}] emission that is much stronger than observed.   However, the model with the best-matching nebular spectrum is more luminous near peak than SN~2019yvq.}\label{fig:compare_model}
\end{center}
\end{figure*}

In this section we examine SN~2019yvq in the context of double-detonation explosions. The double-detonation scenario requires a WD to accrete a surface shell of helium from a binary companion. An ignition in this helium shell can send a shock front into the WD which ignites the C/O core when it converges \citep{Woosley94, Nomoto82}.  The double-detonation mechanism has been considered as a possible channel for some Type Ia SNe, and recently the presence of strong [\ion{Ca}{2}] emission has been pointed to as an identifying signature of these explosions in the nebular phase \citep{Polin20:neb}. 

We compare the event to the explosion models of \citet{Polin19} who use the hydrodynamics code \texttt{Castro} \citep{CASTRO} to simulated double-detonation explosions for a large parameter space of WD and He shell masses. \citet{Polin20:neb} examines these explosion models in the nebular phase using the radiation transport code \texttt{Sedona} \citep{Kasen06:sedona} paired with the NLTE nebular tool \texttt{SedoNeb} \citep{sedoneb} to evolve the homologus ejecta to nebular times while calculating the  gamma-ray  transport of  radioactive decay products. Then \texttt{SedoNeb} is used to calculate the emissivities of each atomic transition by solving for the temperature, ionization  state,  and  NLTE  level  populations. The final step is to integrate this emission to determine the wavelength-dependent flux. 

\citet{Miller20} examined the \citet{Polin19} models to determine the consistency of SN~2019yvq with a double-detonation explosion given the observational properties in the photospheric phase. The best-fit model, a 0.92~\msun\ WD with a 0.04~\msun\ Helium shell (or a total mass of $M_{\rm tot} = 0.96$~\msun), was able to explain most, though not all of the features of SN~2019yvq. Specifically the models showed that a double-detonation can produce the early UV flash exhibited by SN~2019yvq and the best 0.92+0.04 model reproduced the optical brightness during the early flux excess period and at peak brightness. The model, however, struggled to reproduce the velocity evolution of SN~2019yvq exhibiting significantly slower \ion{Si}{2} absorption features than the observed event. 

There is an inherent velocity-luminosity relationship in the 1D double-detonation models of \citet{Polin19}. As a consequence of the WD exploding purely as a detonation the amount of $^{56}$Ni created during core burning is simply a function of the central density (or total mass) of the progenitor. The amount of $^{56}$Ni determines both the peak luminosity of the transient as well as the kinetic energy allowing for the velocity-luminosity relationship to result from a one-parameter function determined by the total mass of the progenitor.

\citet{Polin19} further points to a population of SNe Ia that follow this relationship and a separate group of SNe Ia that have $M_{B} = -19.5$~mag and a peak-brightness \ion{Si}{2} $\lambda 6355$ velocity around -11,000~\kms. This cluster contains most normal SNe Ia, such as SN~2011fe, indicating these are likely not of double-detonation origin. This relationship is, however, based on a set of 1D simulations and has the potential to become more complicated when multi-dimensional effects are introduced.  Furthermore, SN~2019yvq does not follow this relationship, having fast ejecta velocity at peak ($v_{\rm Si~\sc{II}} \approx -15$,000~\kms) that is associated with a high-mass WD, paired with a low luminosity ($M_{g, {\rm ~peak}} \approx -18.5$~mag) that is associated with a low-mass WD. This combination is not just peculiar in the context of a double detonation, but for all SNe~Ia. SN~2019yvq does not lie in the cluster of normal SNe~Ia but rather in a relatively un-populated regime in this parameter space \citep{Miller20}.

Because of the favored photospheric double-denotation model's inability to explain the velocity evolution of SN~2019yvq, we chose to compare our nebular spectrum to the entire suite of \citet{Polin20:neb} models as well as the best-fit model from \citet{Miller20} to independently determine which model best matches the nebular features of SN~2019yvq. \autoref{fig:compare_model} shows the result of this comparison. The best-matching photospheric model, $M_{\rm tot} = 0.96$~\msun, has [\ion{Ca}{2}] $\lambda \lambda$ 7291, 7324 emission that is much stronger than that of SN~2019yvq, while the best-matched nebular model, $M_{tot} = 1.15$~\msun\ (determined by the [\ion{Ca}{2}]/[\ion{Fe}{3}] ratio) is too luminous in the photospheric phase.

The [\ion{Ca}{2}] emission feature is highly sensitive to both the precise amount of Ca produced in the explosion and the distribution of that Ca throughout the ejecta \citep{Polin20:neb}. [\ion{Ca}{2}] is a very efficient cooling line, and tends to dominate the emission features when Ca is co-existent with other coolants. The over production of [\ion{Ca}{2}] in the $M_{\rm tot}=0.96$~\msun\ model could indicate that the 1D double-detonation models distribute too much Ca in the innermost ejecta, allowing for some flux to cool through [\ion{Ca}{2}] when it would otherwise cool through Fe-group elements. However, the $M_{\rm tot}=1.15$~\msun\ model provides a better match to the velocity of SN~2019yvq, exhibiting a \ion{Si}{2} $\lambda$ 6355 of approximately $-14$,500~\kms\ at peak brightness, favoring a higher-mass progenitor for SN~2019yvq.

It is also possible that this discrepancy is due to asymmetries in the explosion and line-of-sight differences not captured in our 1D models. \citet{Townsley19} perform a 2D simulation of the double-detonation of a 1.0~\msun\ WD with 0.02~\msun\ He on its surface. They show that at the time of peak brightness the \ion{Si}{2} velocity is fastest along the pole (in the direction of the initial helium ignition) and slower for viewing angles away from the pole. The bolometric luminosity behaves inversely, such that it is least luminous along the pole and most luminous when viewed from the opposite direction. It is possible that SN~2019yvq is viewed along a line of sight close to the pole, such that it exhibits the rare combination of high \ion{Si}{2} velocity paired with lower luminosity. Future work is necessary to determine how such asymmetries would affect the nebular features of these events. We therefore suggest that while the presence of strong [\ion{Ca}{2}] emission is enough to classify SN~2019yvq as a double-detonation explosion, the exact mass of the progenitor is less certain.

\section{Discussion \& Conclusions}\label{s:disc}

We have gained significant insight about the SN~2019yvq progenitor system and explosion from its nebular spectrum.  There are also broader implications for all SNe~Ia.  SN~2019yvq is another example of a ``normal'' SN~Ia that exhibits an early blue flux excess, but the first with an atypical late-time spectrum.

In almost every case where there is an early flux excess for a SN~Ia, and in all cases where the SN may be considered ``normal,'' the nebular spectra had no obvious peculiarity.  Similarly, the SNe~Ia with peculiar nebular spectra generally lacked evidence of an early flux excess (often because of a lack of data covering the relevant epochs).  The previous exception was the the atypical SN~2002es-like iPTF14atg that had both an early blue flash \citep{Cao15} and [\ion{O}{1}] nebular emission \citep{Kromer16} similar to SN~2010lp \citep{Taubenberger13}.  SN~2019yvq is the first relatively normal SN~Ia with both an early flux excess and a peculiar nebular spectrum.

Notably, none of the ``flux excess'' SNe show evidence for hydrogen or helium emission indicative of swept-up material.  SN~2018fhw had strong H emission at late times, but lacked an early flux excess one might expect from companion interaction \citep{Vallely19}.  SN~2015cp also had strong H emission at very late times ($\sim$650--800~days after peak), but lacks any pre-peak data \citep{Graham19}.  Neither SN~2015cp nor SN~2018fhw had any interaction signatures in their early spectra, unlike SNe~Ia-CSM \citep{Silverman13:csm}.  Some SN~Iax spectra have He emission lines consistent with swept-up material \citep{Foley09:08ha, Foley16:iax, Jacobson-Galan19:iax}, but none have yet had a clear early flux excess.

The most popular progenitor/explosion models for producing excess flux at early times (companion/circumstellar material interaction, surface $^{56}$Ni mixing, double detonation, and violent mergers) have difficulty explaining the nebular spectra of SNe~2017cbv and 2018oh, two normal SNe~Ia with well-observed early flux excess \citep{Hosseinzadeh17:17cbv, Dimitriadis19:18oh_k2, Shappee19}.  In particular, neither SN had detectable [\ion{Ca}{2}] emission, expected for double-detonation explosions \citep{Polin20:neb}. Additionally, SN~2018oh had early blue colors that were inconsistent with double-detonation models \citep{Dimitriadis19:18oh_k2}. Other SNe~Ia that feature early blue colors (but no obvious excess flux) such as SN~2009ig \citep{Foley12:09ig}, SN~2013dy \citep{Zheng13}, and ASASSN-14lp \citep{Shappee16:14lp} also lack evidence of companion interaction or Ca emission in their nebular spectra \citep{Pan15:13dy, Black16, Maguire18, Tucker20:nebular}.

In stark contrast to the other flux-excess SNe, the 7300~\AA\ line complex of SN~2019yvq cannot be explained without strong [\ion{Ca}{2}] emission, a signature of double-detonation explosions (and explicitly outlined by \citet{Miller20} for SN~2019yvq).  All observations of SN~2019yvq, and particularly the early-time flux excess and late-time [\ion{Ca}{2}] emission, are consistent with a thick He shell double-detonation explosion of a sub-Chandrasekhar-mass WD in a binary system.

The double-detonation mechanism requires mass transfer of He onto the primary WD.  Several theoretical studies have indicated that little to no He on the surface of exploding low-mass WDs is needed to reproduce the photospheric properties of normal SNe~Ia \citep{Shen14, Townsley19, Leung20}.  Double detonations with minimal He can be initiated dynamically via an explosion in the accretion stream \citep{Guillochon10}, however, these systems are expected to strip He from the companion WD with masses of $\sim${}$10^{-2}$ to $10^{-1}$~M$_{\sun}$ \citep{Shen17, Tanikawa19}, inconsistent with what is seen for SN~2019yvq.  Alternatively, the He can ignite after a large enough He shell is developed \citep{Shen14}.  \citet{Polin19} showed that minimal-mass shells do not produce early flux excesses like that seen for SN~2019yvq, further excluding a dynamically driven detonation.

SNe~2016hnk and 2018byg are two similar-to-each-other, yet peculiar overall, SNe~Ia that are likely the result of He-shell detonations on the surface of relatively low-mass WDs \citep{De19, Jacobson-Galan20}.  Combined, they had early-time excess flux, strong nebular [\ion{Ca}{2}] emission, and early spectra that demonstrated strong line blanketing from iron-group elements.  \citet{Jacobson-Galan20} modeled the light curves and spectra of SN~2016hnk, finding that the SN was likely the result of a 0.02~M$_{\sun}$ He-shell explosion on the surface of a $0.85~M_{\sun}$ WD.  \citet{De19} estimated that SN~2018byg was produced by the detonation of a massive He shell (0.15$~M_{\sun}$) on a $0.75~M_{\sun}$ WD. These SNe share many features with SN~2019yvq, but the lack of enhanced iron-group elements in the early spectra of SN~2019yvq \citep{Miller20} indicates that SN~2019yvq likely had a significantly larger WD mass than SNe~2016hnk or 2018byg (i.e., $>$0.85~M$_{\sun}$).

\citet{Polin19} also provided evidence that double-detonation SNe originating with varying He-shell masses can be differentiated by their velocity and color. Given the high photospheric velocities, red optical colors, and qualitative similarity to the nebular model of a 1.1~M$_{\sun}$ WD with a 0.05~M$_{\sun}$ He shell (\autoref{fig:compare_model}), we argue that SN~2019yvq was in the distinct thick He shell subclass detailed in \citet{Polin19}. This may also provide evidence that a subset of early ``flux-excess" SNe~Ia are produced by progenitors with thick He shells. There may exist a continuum of thick He shell double-detonation progenitors that ranges from lower-mass events ($<$0.85~\msun) like SNe~2016hnk and 2018byg to higher-mass events ($>$1.1~\msun) like SN~2019yvq. Furthermore, given the likely [\ion{Ca}{2}] presence in some fast-declining SNe~Ia \citep[e.g., SN~1999by;][]{Blondin18}, it is reasonable to expect that they may be produced by a double-detonation progenitor. However, since these SNe do not show prominent early excess flux, they are likely not produced through the thick He shell channel.

Assuming that the progenitor system of SN~2019yvq is unique compared to the population of normal SNe~Ia with nebular spectra, and SN~2019yvq is consistent with a thick He shell double-detonation explosion, we can estimate the fraction of normal SNe~Ia that arise from this progenitor channel. The nebular spectrum of SN~2019yvq was acquired +153~days after peak brightness and SN~2019yvq had $M_{g} = -14.0 \pm 0.1$~mag at this time. Given a typical nebular spectroscopy survey limiting magnitude of 21.5~mag, the nebular phase spectrum of SN~2019yvq would have been detectable to 124~Mpc. The comprehensive nebular sample provided by \citet{Tucker20:nebular} contains 94 normal SNe~Ia within this volume. Using Poisson statistics, we determine that the fraction of normal SNe~Ia that are SN~2019yvq-like double-detonation SNe is $1.1^{+2.1}_{-1.1}$\% (90th-percentile confidence range).

The simulations in \citet{Shen18a} favor a $\sim$1.0~M$_{\sun}$ progenitor for typical SN~2011fe-like SNe~Ia. Since SNe~Ia typically do not show strong [\ion{Ca}{2}] emission, they must either originate from a channel that does not have a double-detonation explosion or must come from WDs more massive than the progenitor of SN~2019yvq (i.e., $>$1.1~M$_{\sun}$). This presents a problem for minimal He mass double-detonation explosions as the dominant path to creating normal SNe~Ia since $>$1.1~M$_{\sun}$ WDs are rare \citep{Kilic18}. While the WDs may be born at a lower mass and accrete to a higher mass, reaching this higher mass can still be difficult, especially if the accretion is from a low-mass He WD.  Furthermore, their synthetic spectra in the photospheric phase of massive WD explosions tend to generate higher velocities than observed in normal SNe~Ia. Surviving WD companions of double-degenerate systems have been detected \citep{Shen18}, but it is still uncertain whether the implied rate of these progenitors can account for the majority of SNe~Ia.

\citet{Shen18} used {\it Gaia} parallaxes and proper motions to search for hyper-velocity stars that could be the surviving companion star from a double-detonation progenitor system. They estimated that if all SNe in the Milky Way originated from the dynamically driven double-degenerate double-detonation (D$^6$) channel, they would detect 22 runaway WDs within 1 kpc of the Sun in the {\it Gaia} DR2 sample. \citet{Shen18} found three likely runaway WDs, however, these ranged from a distance of 1.0 to 2.3 kpc from the Sun and were only detectable because of their higher luminosity than normal WDs. This was an incredible success of the theory, but the detection rate, when considering the higher luminosity, is consistent with only 1.1\% of SNe~Ia producing runaway WDs.  Assuming Poisson statistics, we determine that 95\% confidence interval of the observed-to-predicted rate is $0.35-3.0$\%.  While there are several selection effects that we ignore for both measurements, we note that the rate of SN~2019yvq-like events is consistent with the rate of hyper-velocity white dwarfs in the Milky Way. Since the rates are similar, it is possible that all double-detonation explosions (those with stable and unstable mass transfer) account for only a fraction of normal SNe~Ia, with an additional channel possibly necessary to produce the bulk of normal SNe~Ia. 

The diversity of SNe~Ia in the nebular phase, and particularly the tell-tale signs of different progenitor/explosion scenarios for SNe~2010lp, 2018fhw, and 2019yvq, point to a variety of paths to have SNe~Ia with similar near-peak observables.  These SNe provide some of the strongest support for violent-merger, single-degenerate, and double-detonation models, respectively. Yet the rarity of these kinds of SNe and the divergence from the majority of SNe~Ia suggests that these channels are either not the dominant channels producing most SNe~Ia, or these examples are extrema of the most-common channel.

While some of these rare SNe~Ia would likely be excluded from cosmological samples, SN~2019yvq is not clearly an outlier.  Although its decline rate is faster than the average SN~Ia \citep{Miller20}, it is not large enough to be clearly rejected, especially for lower signal-to-noise ratio or more sparsely covered light curves.  Additionally, its red color and low peak luminosity are consistent with its decline rate.  Future detailed simulations will reveal if SN~2019yvq-like SNe impact cosmological measurements.

We summarize our analysis of the SN~2019yvq nebular spectrum below:

\begin{itemize}
    \item The +153-day nebular spectrum of SN~2019yvq exhibits strong [\ion{Ca}{2}] $\lambda\lambda$7291, 7324 and Ca NIR triplet emission features. The nebular spectra of some other fast-declining SNe~Ia likely have contributions from [\ion{Ca}{2}] $\lambda\lambda$7291, 7324 emission, but the relative strengths of the [\ion{Ca}{2}] to [\ion{Fe}{2}] and [\ion{Ni}{2}] in these SNe are more difficult to determine. In some more extreme SNe~Ia such as SN~1999by, [\ion{Ca}{2}] likely dominates in this region, but the narrow Fe and Co features in other region of the spectra are inconsistent with SN~2019yvq.
    \item We fit a multiple-component Gaussian emission model to the 7300~\AA\ line complex consisting of [\ion{Fe}{2}], [\ion{Ni}{2}], and [\ion{Ca}{2}] emission, finding that all components are blueshifted relative to their rest-frame wavelengths. Blueshifted nebular lines are atypical for high-velocity SNe~Ia such as SN~2019yvq. 
    \item We find no evidence for swept-up material in the nebular spectrum of SN~2019yvq. Our limits on the amount of hydrogen and helium mass are $<${}$2.8 \times 10^{-4}$ and $2.4 \times 10^{-4}$~M$_{\sun}$, respectively.
    \item We also do not detect [\ion{O}{1}] $\lambda\lambda$6300, 6364 emission, an expected feature if there is significant unburned material in the ejecta.
    \item A comparison to the double-detonation models from \citet{Polin20:neb} reveals that SN~2019yvq was likely the result of double-detonation explosion. The strength of the Ca emission indicates a larger progenitor mass ($1.15~M_{\odot}$), however, a lower progenitor mass still better reproduces the early light-curve and low peak luminosity.
    \item The rarity of SN~2019yvq-like events suggests that thick He shell double-detonations make up $1.1^{+2.1}_{-1.1}\%$ of the normal SN~Ia population.
\end{itemize}

Continued observations of SN~2019yvq will further enhance this picture with future observations potentially revealing additional insight into the progenitor system and explosion.  Continued monitoring of the [\ion{Ca}{2}] emission will allow models to better separate abundance, ionization, and asymmetry.  Additional data such as spectropolarimetry of similar events will be especially valuable to untangle the early-time emission.

More photometric observations of SNe~Ia in their infancy are needed to better understand the population with early excess flux. Several subclasses of SNe~Ia have both early excess flux and spectral signatures in the nebular phase that indicate a variety of progenitor channels \citep{Taubenberger13, Cao15, Kromer16, De19, Jacobson-Galan19}.  More typical SNe~Ia with early excess flux lack clear late-time signatures \citep{Hosseinzadeh17:17cbv, Sand18, Dimitriadis19:18oh_k2, Dimitriadis19:18oh_neb, Shappee19, Tucker19}.  And some SNe~Ia with peculiar nebular spectra \citep{Taubenberger13:10lp, Kollmeier19} do not have detected early-time excess flux, often to deep limits.  High-cadence surveys of the local volume where one can hope to obtain a nebular spectrum will be critical.

\acknowledgments
M.R.S.\ is supported by the National Science Foundation Graduate Research Fellowship Program Under Grant No.\ 1842400. The UCSC team is supported in part by NASA grant NNG17PX03C, NSF grant AST-1815935, the Gordon \& Betty Moore Foundation, the Heising-Simons Foundation, and by a fellowship from the David and Lucile Packard Foundation to R.J.F. The Computational HEP program in The Department of Energy's Science Office of High Energy Physics provided resources through Grant \#KA2401022. Calculations presented in this paper used resources of the National Energy Research Scientific Computing Center (NERSC), which is supported by the Office of Science of the U.S. Department of Energy under Contract No.\ DE-AC02-05CH11231.

The data presented herein were obtained at the W.\ M.\ Keck Observatory, which is operated as a scientific partnership among the California Institute of Technology, the University of California and the National Aeronautics and Space Administration. The Observatory was made possible by the generous financial support of the W.\ M.\ Keck Foundation. We thank Elena Manjavacas and Lucas Fuhrman for assistance with these observations. The authors wish to recognize and acknowledge the very significant cultural role and reverence that the summit of Maunakea has always had within the indigenous Hawaiian community.  We are most fortunate to have the opportunity to conduct observations from this mountain.
\vspace{5mm}
\facilities{Keck:I (LRIS)}

\software{astropy \citep{astropy}, kaepora \citep{Siebert19}, Castro \citep{CASTRO}, Sedona \citep{Kasen06:sedona}, SedoNeb \citep{sedoneb}}

\bibliography{astro_refs}
\bibliographystyle{aasjournal}

\end{document}